\documentstyle[12pt]{article}
        \catcode`\@=11
        \@addtoreset{equation}{section}

\newcommand{\al}[1]{
\begin{eqnarray}#1
\end{eqnarray}}
\newcommand{\eq}[1]{\begin{equation}#1\end{equation}}

\begin{document}
\makeatletter
\def\siml{\mathrel{\mathpalette\gl@align<}}
\def\simg{\mathrel{\mathpalette\gl@align>}}
\def\gl@align#1#2{\lower.6ex\vbox{\baselineskip\z@skip\lineskip\z@
 \ialign{$\m@th#1\hfill##\hfil$\crcr#2\crcr{\sim}\crcr}}}
\makeatother
\hbadness=10000
\hbadness=10000
\begin{titlepage}
\nopagebreak
\def\thefootnote{\fnsymbol{footnote}}
\begin{flushright}

        {\normalsize
 LMU-TPW 95-2\\
March, 1995   }\\
\end{flushright}
\begin{center}
\renewcommand{\thefootnote}{\fnsymbol{footnote}}
{\large \bf Implications of Non-universal Soft Masses
on Gauge Coupling Unification}


{\bf Richard Altendorfer
\footnote[1]{e-mail: richard@lswes8.ls-wess.physik.uni-muenchen.de}
and
Tatsuo Kobayashi
\footnote[2]{Alexander von Humboldt Fellow \\
\phantom{xxx}e-mail: kobayash@lswes8.ls-wess.physik.uni-muenchen.de}
}

       Sektion Physik, Universit\"at M\"unchen, \\

       Theresienstr. 37, D-80333 M\"unchen, Germany \\

\end{center}

\nopagebreak
\begin{abstract}
We study the gauge coupling unification
of the minimal supersymmetric standard model with non-universal soft scalar
and gaugino masses.
The unification scale of the gauge couplings is estimated for
non-universal cases.
It is sensitive to the non-universality.
It turns out that these cases can be combined with the assumption of string
unification, which leads to a prediction of $\sin^2 \theta_W(M_Z)$ and $k_1$,
the normalisation of the $U(1)_Y$ generator.
String unification predicts $k_1=1.3 - 1.4$.
These values have non-trivial implications on string model building.
Two-loop corrections are also calculated.
Some of these cases exhibit a large discrepancy between experiment and
 string unification.
We calculate string threshold corrections to explain the discrepancy.

\end{abstract}

\vfill
\end{titlepage}
\pagestyle{plain}
\newpage
\voffset = 0.5 cm

\vspace{0.8 cm}
\leftline{\large \bf 1. Introduction}
\vspace{0.8 cm}

Supersymmetric models are some of the most promising extensions of the standard
model (SM) \cite{Nilles}. For a phenomenologically viable scenario,
supersymmetry (SUSY) is broken by soft SUSY breaking terms,
scalar masses, gaugino masses and trilinear and bilinear couplings of
scalar fields.
These soft terms have implications on low energy physics.

Aspects of the soft terms can be derived from underlying supergravity
theories or superstring theories.
In most of the studies one assumes universality of soft scalar masses and
gaugino masses at the unification scale $M_X$.
However breaking of supergravity in general leads to non-universal
soft scalar masses and gaugino masses \cite{Ibanez,KL,Brignole}.
Further it was shown in ref.\cite{KSYY} how
cases where soft scalar masses are fairly different from each other can be
obtained.
Therefore it is important to study SUSY models with non-universal soft
masses and to show which phenomenological features are sensitive or
insensitive to such non-universality.

Effects of the non-universality have been studied recently for some
phenomenological aspects, e.g. gauge coupling unification, Yukawa
coupling unification, radiative symmetry breaking and
neutron electric dipole moment \cite{KSY,Non}.
Some types of non-universality are constrained by flavor changing neutral
currents (FCNC) \cite{FCNC}.
The gauge couplings of the SM groups are unified at
 $M_X \approx 3 \times 10^{16}$GeV within the framework of the minimal
supersymmetric standard model (MSSM) with universal soft masses
\cite{MX}.
In ref.\cite{KSY} it was shown that the gauge coupling unification of $SU(3)$
and $SU(2)$ is sensitive to the non-universality of the soft scalar masses.
In most of the non-universal cases, the unification scale $M_X$ rises
higher than $3 \times 10^{16}$GeV.
In some cases the unification scale $M_X$ becomes higher than
 $10^{17}$GeV and closer to a string scale
 $M_{st}=5.27\times g_{st} \times 10^{17}$GeV \cite{Kaplunovsky}
with $g_{st}$ being a universal string coupling of order one.

One of the strongest arguments in favour of supersymmetric
grand unified theories (GUTs) is the unification of the
three SM interactions at $M_X \approx 3 \times 10^{16}$GeV and the correct
prediction
of $\sin^2 \theta_W(M_Z)$ within the experimental bounds \cite{MX}.
Unfortunately, not all features of
GUTs are theoretically satisfactory, e.g. the doublet-triplet splitting
problem.
String theories realize another class of unified theories.
One does not need a unified group like $SU(5)$ or $SO(10)$.
Each gauge coupling $g_a$ is related to $g_{st}$ as
 $k_ag_a^2=g_{st}^2$ at $M_{st}$ \cite{Gi}, where $k_a$ is
a Kac-Moody level of each gauge group.
Hence a direct string unification of the MSSM is free from
the theoretical problems of SUSY-GUTs.
For levels, we take $k_a=1$ for $SU(3)$ and $SU(2)$ \footnote[1]{
Non-abelian Gauge groups with $k_a \neq 1$ are discussed in
refs.\cite{Fo,highk}.}.
The normalization of the $U(1)_Y$ generator $k_1$ is treated as a free
parameter \cite{Ib}.

The purpose of this paper is to study the evolution of the gauge couplings in
the MSSM for non-universal cases and to consider their implications from
the viewpoint of string theories.
Here we discuss non-universal cases where gaugino masses as well as soft
scalar masses are non-universal.
The non-universality of the gaugino masses entails non-universality of
the soft scalar masses by radiative corrections from $M_X$ or $M_{st}$ to
the weak scale $M_Z$.
To obtain a non-universal case at $M_X$ might require `fine tuning' of
parameters like goldstino angles.
However, it is important to investigate all possibilities at present.
We study the running of the gauge couplings in two ways.
One is to run the renormalisation group equations (RGEs) of the gauge
couplings bottom-up using experimental data and then to estimate the
unification
scale in each non-universal case.
The other is to run the RGEs top-down assuming string unification and
to compare with the experimental data.
Results obtained in both ways are considered from the viewpoint of string
theories, especiallly orbifold models \cite{Orbi}.

This paper is organized as follows.
In section 2 we review in brief soft masses derived from supergravity
theories.
In subsection 3.1 we estimate unification scales of $SU(3)$ and $SU(2)$
in non-universal cases using the experimental values of $\alpha$,
$\sin \theta_W$ and $\alpha_3$ and one-loop RGEs of the gauge couplings.
Radiative corrections to soft masses are calculated.
In subsection 3.2 we assume string unification and run one-loop RGEs
from the string scale down to $M_Z$.
Two-loop corrections are also computed.
In section 4, we discuss results obtained in section 3 from the viewpoint
of string theories.
In subsection 4.1 we investigate threshold corrections due to
massive string modes which are needed to explain the difference between
experiment
and the prediction by string unification.
Implications of $k_1$ are discussed in subsection 4.2.
Section 5 is devoted to conclusions and discussions.

\vspace{0.8 cm}
\leftline{\large \bf 2. Soft Masses}
\vspace{0.8 cm}

In this section we review soft scalar masses and gaugino masses
derived from supergravity theories \cite{KL}.
A supergravity Lagrangian is characterized by a K\"ahler potential $K$,
a superpotential $W$ and a gauge kinetic function $f_a$,
where the subscript $a$ represents a gauge group.
Here we assume that fields $\Phi^m$ have a non-perturbative potential
$\widehat W(\Phi )$ leading to SUSY breaking.
The K\"ahler potential and the total superpotential are expressed as
follows
$$K=\kappa^{-2}\widehat K(\Phi,\bar \Phi)+K(\Phi,\bar \Phi)_{I \bar J}
Q^I\bar Q^{\bar J} +({1 \over 2}H(\Phi,\bar \Phi)_{I J} Q^I Q^J +
{\rm h.c.})+ \cdots,$$
$$W=\widehat W(\Phi) + \tilde W(\Phi,Q),
\eqno(2.1)$$
where $\kappa^2=8\pi/M^2_{pl}$ and $Q^I$ are chiral superfields.
The dots stand for terms of higher orders in $Q^I$.
We have a scalar potential $V$ as follows,
$$V=\kappa^{-2}e^G[G_\alpha(G^{-1})^{\alpha \bar \beta}G_{\bar \beta}
-3\kappa^{-2}],
\eqno(2.2)
$$ where $G=K+\kappa^{-2}\log \kappa^6 |W|^2$ and the indices $\alpha$
and $\beta$ represent $Q^I$ and $\Phi^m$.
Here we do not consider the D-term contribution to $V$.
The gravitino mass $m_{3/2}$ is obtained as
$$ m_{3/2}=\kappa^2 e^{\widehat K/2}|\widehat W|.
\eqno(2.3)$$
In (2.2) we take the flat limit ($M_{pl} \rightarrow \infty$) while keeping
$m_{3/2}$ fixed.
Then we obtain soft scalar masses $m_{I \bar J}$ for canonically normalized
fields $Q^I$ as follows,
$$m_{I \bar J}^2K_{I \bar J}=m_{3/2}^2K_{I \bar J}-
F^mF^{\bar n}[\partial_m \partial_{\bar n}K_{I \bar J}-(\partial_{\bar n}
K_{K \bar J})K^{K \bar L}(\partial_mK_{I \bar L})] +
\kappa^2V_0K_{I \bar J},
\eqno(2.4)$$
where $F^m$ are F-terms of $\Phi^m$, $\partial_m$ represent derivatives with
respect to
$\Phi^m$ and $V_0$ is the cosmological constant, which is expressed as
$$V_0=\kappa^{-2}(F^mF^{\bar n}\partial_m \partial_{\bar n}
\widehat K-3m_{3/2}^2).
\eqno(2.5)$$
Further we obtain gaugino masses $M_a$ as
$$M_a=F^m\partial_m \ln Re f_a.
\eqno(2.6)$$

In general the form of the K\"ahler metric $K_{I \bar J}$ depends on each
field $Q^I$ and the gauge kinetic term $f_a$ is
also dependent on the gauge group.
Thus we have non-universal soft scalar masses and gaugino masses.
Therefore it is important to study the effects of the non-universality on
phenomenological aspects.

For example orbifold models have the following K\"ahler potential
\cite{Kahler}:
$$K=-\log (S+\bar S)-\sum_i \log (T^i+\bar T^i)
+\prod_i (T^i+\bar T^i)^{n^i_I}Q^I{\bar Q}^{\bar I},
\eqno(2.7)$$
where $S$ is the dilaton field, $T^i$ are moduli fields and
$n^i_I$ are modular weights of $Q^I$ corresponding to $T^i$.
At tree level the gauge kinetic function is obtained as $f_a=k_aS$ and
it is independent of gauge groups.
However one-loop corrections induce $T$-dependent threshold corrections,
which depend on the gauge groups \cite{Thres}.
It is plausible that $S$ and $T^i$ contribute to the non-perturbative
superpotential $\widehat W$ which breaks SUSY.
If $S$ contributes dominantly to the SUSY breaking, we obtain universal
soft scalar and gaugino masses \cite{BLM}.
Otherwise the soft scalar masses depend on their modular weights and
$T$-dependent threshold corrections lead to non-universal gaugino masses.
Non-universality of soft scalar masses was discussed in
refs.\cite{Brignole,KSYY} under certain assumptions about $\widehat W$.
In ref.\cite{Carlos} a concrete superpotential induced by gaugino
condensation was used to derive soft scalar masses which are universal
and heavier than a universal gaugino mass.

\vspace{0.8 cm}
\leftline{\large \bf 3. Running of gauge couplings}
\leftline{\large \bf 3.1 Unification scale}
\vspace{0.8 cm}

In this section we study effects of non-universal masses on
the evolution of the gauge couplings within the framework of the MSSM.
First of all we classify non-universalities of soft scalar masses
and gaugino masses.
For simplicity, we divide sfermions and gauginos into two groups,
${\cal A}$ and ${\cal B}$.
We assume that the superpartners in Group ${\cal A}$ have a representative
mass $M_S$ ($M_S \geq M_Z$) and the superpartners in Group ${\cal B}$
have a mass $M_Z$ at the weak scale.
Note that soft scalar masses get radiative corrections of order of
gaugino masses at low energies.
For example squarks have at least a mass of order of a gluino mass
at low energies. Thus we do not consider here the case where $\lambda_3$
belongs to ${\cal A}$ while squarks belong to ${\cal B}$.
We consider the typical cases shown in Table 1,
where $\lambda_a$ represent $SU(a)$ gauginos and $Q$, $U$, $D$ and $L$ denote
squark doublets, up- and down-squark singlets and slepton doublets
respectively.
Case~I corresponds to the ordinary MSSM with the SUSY breaking scale $M_S$.
We assume slepton singlets $E$ and the $U(1)_Y$ gaugino $\lambda_1$ belonging
to Group ${\cal B}$ in each case except in Case~I where all
superpartners belong to Group ${\cal A}$.
Here the generation indices are abbreviated, because we assume degeneracies
between the generations in order to avoid FCNC.
Further, we assume that the Higgs sector below $M_S$ has the same structure
as the SM.
In Cases III -- VI, soft scalar masses are non-universal.
On top of that, gaugino masses are also non-universal in Cases A -- D.

Now we study the unification scale of $SU(3)$ and $SU(2)$ gauge couplings
in each non-universal case.
Note that the gauge coupling of $U(1)_Y$ can always be unified with the
other couplings using a suitable value of $k_1$.
We decouple matter fields of Group ${\cal A}$ at $M_S$ in the RGEs of the
gauge couplings.
Then the gauge coupling constant evolves at $\mu$ ($\mu>M_S$) as follows
$$\alpha_a^{-1}(\mu)=\alpha_a^{-1}(M_Z)-{\bar b_a \over 2 \pi} \ln
{M_S \over M_Z}-{b_a \over 2 \pi} \ln {\mu \over M_S},
\eqno(3.1)$$
where $b_a$ are the one-loop $\beta$-function coefficients of the MSSM,
i.e. $(b_1,b_2,b_3)=(11,1,-3)$.
Note that we take $\alpha_1=g_1^2/4\pi$, because in our approach $k_1$ is
a free parameter.
In (3.1), $\bar b_a$ denote one-loop $\beta$-function coefficients  between
$M_S$ and $M_Z$.
The values of $(\bar b_1,\bar b_2,\bar b_3)$ are obtained for each case as
shown in the fourth column of Table 1.

Taking $\alpha_X=\alpha_3(M_X)=\alpha_2(M_X)$ in (3.1), we obtain the
unification scale $M_X$ of the $SU(3)$ and $SU(2)$ gauge couplings as
follows,
$$\ln M_X={\bar b_2- \bar b_3 \over b_3-b_2}\ln {M_S \over M_Z} +
\ln M_S + {2 \pi \over b_2-b_3}(\alpha_2^{-1}(M_Z)-\alpha_3^{-1}(M_Z)).
\eqno(3.2)$$
The smaller the value of $(\bar b_2 - \bar b_3)$ the higher is the
unification scale $M_X$, because $b_3-b_2=-4$.
Here we estimate the unification scales $M_X$ in non-universal cases using
$M_Z=91.173$GeV, $\alpha^{-1}(M_Z)=127.9$, $\sin^2\theta_W(M_Z)=0.2321$
 and $\alpha_3(M_Z)=0.118$.
For Cases I and A -- D, the unification scale $M_X$ is shown against
 $M_S$ in Figure 1, where x and y denote $\log_{10}M_S$ and
 $\log_{10}M_X$ (in GeV), respectively.
In ref.\cite{KSY}, a similar figure on $M_X$ was shown for
Case~I -- VI.
For $M_S=1$TeV, the unification scales $M_X$(in GeV) of all the
cases are also found in the second column of Table 2, where the numbers
in the parentheses represent the corresponding values of the case with
$M_S=\sqrt {10}$GeV.
In Case~I, $M_X$ is stable against $M_S$.
Because of the value $\bar b_2-\bar b_3= 23/6$ being very close to
the value of $b_2-b_3=4$ the $M_S$ dependence in (3.2) is suppressed.
The unification scale is very sensitive to the non-universality of
the gaugino masses.
$SU(2)$ gauginos belonging to group ${\cal A}$
lead to a higher unification scale, while
gluinos belonging to group ${\cal A}$ lead to a lower value.
Most of the cases II -- VI result in a higher unification scale than Case~I.
Note that the estimation of $M_X$ includes an uncertainty of order
$10^{0.3}$GeV as usual \cite{MX}.
Suppose that $\alpha_{st}=g_{st}^2/4 \pi =1/25$, then we have
$M_{st}=3.7 \times 10^{17}$GeV.
If we take $M_S=2.5$TeV for Case~A,
the unification scale $M_X$ becomes $3.7 \times 10^{17}$GeV.
Case~III with $M_S=7.1$TeV and Case~B with $M_S=3.6$TeV lead to
 $M_X=10^{17.3}$GeV, which coincides with $M_{st}$ within the uncertainty
of $10^{0.3}$GeV.

Furthermore the third column of Table 2 shows values of the unified coupling
$\alpha_X^{-1}$.
In all cases the value of $\alpha_X^{-1}$ increases as $M_S$ becomes
higher.
Similarly we can run the $U(1)_Y$ gauge coupling $\alpha_1$ from $M_Z$
to $M_X$.
The fourth column of Table 2 shows the ratios $\alpha_X/\alpha_1(M_X)$,
which correspond to the values of $k_1$.
The value $k_1$ is also sensitive to the non-universality.
It takes values from 1.4 to 1.7 in the case with $M_S=1$TeV.
In some cases, e.g., Case~V, VI, D, the value is stable around 1.6
against $M_S$.
This value corresponds to the GUT prediction $k_1=5/3$.
Even if $M_S=10$TeV, Case~V, VI and D lead to
$\alpha_X/\alpha_1(M_X)=1.58$, 1.64 and 1.67 respectively.
It seems that a higher unification scale corresponds to a smaller
value of $k_1$.
In all cases with $M_S=M_Z$, we obtain $M_X=10^{16.4}$GeV,
$\alpha_X^{-1}=24.4$ and $\alpha_X/\alpha_1(M_X)=1.63$.

Our classification is based on the fact that by radiative corrections
squarks are at least as heavy as gluinos and SU(2) doublet sfermions
are at least as heavy as SU(2) gauginos, even if
these sfermions are massless at $M_X$.
Here we investigate this assumption by calculating these radiative
corrections.
We use the following RGEs,
$${dM_a \over dt}={\widetilde b_a \over 2\pi}\alpha_a M_a,
\eqno(3.3)$$
$${dm_I^2 \over dt}={1 \over 2\pi}\left(-4\sum_a C_a(\underline R_I)
M_a^2\alpha_a +({\rm Yukawa\ terms})\right),
\eqno(3.4)$$
where $t=\ln \mu$ and $C_a(\underline R_I)$ is the quadratic Casimir
of the representation $\underline R_I$ corresponding to each scalar field.
We neglect the contributions due to the Yukawa terms in (3.4).
If SUSY is a good symmetry, then $\widetilde b_a=b_a$.
Thus we have the relation $M_a(\mu)/\alpha_a(\mu)=$ constant for $\mu > M_S$.
If sfermions decouple at $M_S$, $\widetilde b_a$ is not always equal to
$\bar b_a$ for energies below $M_S$.
This is because one-loop corrections to the gaugino masses include graphs
which have fermions and their superpartners simultaneously
in internal lines.\footnote[2]
{Some SUSY relations between several couplings are broken
below the SUSY threshold.
For example, gaugino-scalar-fermion couplings are independent of
the corresponding gauge couplings below $M_S$.
That leads to further corrections \cite{Chankowski}.
We neglect here such corrections.}
Therefore we obtain the following relation below $M_S$ for $\mu < M_S$,
$$(M_a(\mu))^{\bar b_a}(\alpha_a(\mu))^{-\widetilde b_a}={\rm constant}.
\eqno(3.5)$$
We have $(\widetilde b_1,\widetilde b_3)=(6,-6)$ for Case~A,
$(\widetilde b_1,\widetilde b_3)=(3,-9)$ for Case~B,
$(\widetilde b_1,\widetilde b_2)=(9/2,-9/2)$ for Case~C and
$(\widetilde b_1,\widetilde b_2)=(3,-6)$ for Case~D.

Using eqs. $(3.3 - 3.5)$ we calculate radiative corrections to the gaugino
masses
 $M_a$ and the soft scalar masses $m_I$ for Case~A -- D.
Tables 3 and 4 show the results in the case with $M_S=1$TeV.
The second row of Table 3 lists ratios of $M_3(\mu)$ to $M_3(M_X)$,
where $\mu=M_Z$ for Case~A and B, and $\mu=M_S$ for Case~C and D.
Similarly corrections to $M_2$ and $M_1$ are found in the other rows.
The second column shows these values for Case~I for comparison.
Table 4 lists corrections to the scalar masses
$\Delta m^2_I \equiv m^2_I(m_I)-m^2(M_X)$.
It is clear from (3.4) that $\Delta m^2_I$ is represented by a linear
combination of $M_3^2(M_X)$, $M_2^2(M_X)$ and $M_1^2(M_X)$.
Their coefficients of the linear combinations are found in Table 4.
For example the third row for Case~A represents
$\Delta m^2_Q=m^2_Q(M_S)-m^2_Q(M_X)$ as
$$\Delta m_Q^2=5.0 \times M_3^2(M_X) + 0.46 \times M_2^2(M_X) +
4.3 \times 10^{-3} \times M_1^2(M_X).
\eqno(3.6)$$
In Table 4 values of $\Delta m_D^2$ are omitted, because they are equal
to the corresponding values of $\Delta m_{U}^2$ except small contributions
due to $M_1$.
The values in the parentheses of Table 3 and 4 represent the corresponding
values of the case where $M_S=\sqrt {10}$TeV.
For example the squark masses get the corrections $\Delta m_{Q,U,D}^2
\approx 3.8 \times M_3^2(M_X)$ in Case~C or D with $M_S=1$TeV.
Thus even though the squarks are massless at $M_X$, they have
$m=1.9 \times M_3(M_X)$ at 1TeV.
On the other hand, the scalar fields $L$ get the correction
$\Delta m_L^2 \approx 0.46 \times M_2^2(M_X)$ in Case~A or B
with $M_S=1$TeV.
Even though $m_L$ vanishes at $M_X$, the scalar field $L$ acquires
$m_L=0.68 \times M_2(M_X)=0.82 \times M_2(1{\rm TeV})$ at 1TeV.
To obtain the second relation, we use $M_2({\rm 1TeV})/M_2(M_X)=0.83$ for
Case~A and B shown in Table 3.
Hence the scalar field $L$ acquires a mass of
order of $M_2(M_X)$ or $M_2$(1TeV) through radiative corrections.

In the above, we have assumed that the masses of $E$ and $\lambda_1$ are of
order of $M_Z$.
Even if we alter this assumption, its effects on the results of Tables 3 and
4 can be neglected.
For example we consider Case~A where the masses of $E$ and
$\lambda_1$ are of order of $M_S$.
For $M_S=1$TeV, we have $M_1(M_S)/M_1(M_X)=0.39$ and the coefficients of
$M_1^2(M_X)$ for $\Delta m^2_U(=m^2_U(M_Z)-m^2_U(M_X))$,
$\Delta m^2_L(=m^2_L(M_S)-m^2_L(M_X))$ and
$\Delta m^2_E(=m^2_E(M_Z)-m^2_E(M_X))$ become $6.8 \times 10^{-2}$,
$3.8 \times 10^{-2}$ and $1.5 \times 10^{-1}$ respectively.
The other values of Tables 3 and 4 are not changed.

\vspace{0.8 cm}
\leftline{\large \bf 3.2 String unification}
\vspace{0.8 cm}
\setcounter{section}{3}
\setcounter{equation}{6}

In SUSY-GUTs the normalisation $k_1$ of the $U(1)_Y$ generator is determined
by the requirement that the
 $U(1)_Y$ generator belongs to the set of generators of the unification gauge
group: $k_1 = 5/3$.
A free parameter of the theory is $M_{X}$, which can only be calculated with
the
knowledge (i.e.
measurement) of the low energy gauge couplings, preferentially at $M_Z$. In
string unification
scenarios, the situation is different. The gauge coupling constants $g_a$ of
the
three SM
interactions
are related  at the string scale $M_X=M_{st}$ by \cite{Gi}:
\eq{
k_3g_3^2 = k_2g_2^2 = k_1g_1^2,
}
where we take $k_2 = k_3 =1$.
This boundary condition is independent of any GUT type unification of strong
and
weak interactions
and
is a proper consequence of string unification. On the other hand there is
little
model
independent information known about $k_1$, except that it should be a rational
number with
\cite{Fo}:
\eq{
k_1 \geq 1. \label{Constraint}
}
The lepton singlet field $E$ is not allowed in models with $k_1<1$.
As we will constrain ourselves to a model independent discussion, $k_1$ can
therefore be
regarded as a
free parameter of the theory. Nevertheless, as shown in ref.\cite{Ib}, the
predictivity of
direct string unification is not smaller than in GUT unification, because in
the
string case
the incorporation of gravity leads to another constraint yielding a prediction
of the unification
scale \cite{Kaplunovsky}: $M_{st} = 5.27 \times g_{st}\times 10^{17}$GeV.
the
experimental data
At one-loop level, the running coupling constants with one intermediate
breaking
scale evolve
according to the following equations:
\eq{
\alpha_a^{-1}(M_Z)={k_a\over \alpha_{st}} +{b_a\over 2\pi}\ln ({M_{st}\over
M_S})+
{\bar b_a\over 2\pi}\ln ({M_S\over M_Z}), \label{GQW}
}
with the same notation adopted as before. After elimination of $k_1$ we obtain:
\al{
\sin^2\theta_W(M_Z)&=&{\alpha(M_Z)\over \alpha_2(M_Z)} \label{Wein} \\
&=&\alpha(M_Z)({4\pi\over g_{st}^2} + {b_2\over 2\pi} \ln
g_{st} + {b_2\over 2\pi}\ln ({5.27\times 10^{17}\over M_S})
+ {\bar b_2\over 2\pi}\ln ({M_S\over M_Z})). \nonumber
}

Using (\ref{GQW}) for $\alpha_3(M_Z)$, we can compute $g_{st}$ in (\ref{Wein})
and then
obtain a constraint in the $\sin^2\theta_W(M_Z)$ - $\alpha_3(M_Z)$ plane.
This constraint is plotted in Figure 2 for all of the cases with $M_S=1$TeV.
In this figure, a and sin denote $\alpha_3(M_Z)$ and $\sin^2\theta_W(M_Z)$
respectively and
the experimental values for
 $\sin^2\theta_W(M_Z)=0.2321 \pm 0.004$ and
 $\alpha_3(M_Z)=0.118 \pm 0.007$ are also displayed.
Contrary to ref. \cite{Ib} we allow for different values of $g_{st}$ in
$M_{st}$.
This gain of exactitude in  $M_{st}$ causing changes of $\Delta M_{st}\approx
-0.1
\times 10^{17}$GeV
is however of the order of the uncertainty of the numerical factor 5.27 in
$M_{st}$
\cite{Kaplunovsky}.

Similarly we can get a relation between $k_1$ and $\sin^2\theta_W(M_Z)$
or $k_1$ and $\alpha_3(M_Z)$, using (3.9) for $\alpha_1$.
Further the second and third columns of Table 5 show the values
 $\sin^2\theta_W(M_Z)$ and $\alpha_3(M_Z)$ of the nearest point of each
curve to the center of the error cross.
Corresponding values of $k_1$ are also found in the fourth column of
Table 5.
Case I -- VI are fairly equal in their prediction of the Weinberg angle,
although Case II, III and IV do a little bit better.
Case I, V and VI are almost degenerate.
Among the cases classified above Case A and B do exceptionally
well, whereas Case C and D seem to be ruled out by this calculation.
In addition all of the cases predict $k_1=1.3 - 1.4$, which is
consistent with (\ref{Constraint}) and agrees with the result of ref.\cite{Ib}.

As an example, the qualitative features of the dependence on $M_S$ is shown
in Figure 3 for Case I, II and A with some values of $M_S$.
As expected from the discussion of the dependence of the unification scale on
$M_S$,
the curve for Case I is not very sensitive to a shift of $M_S$ from 1TeV to
10TeV,
whereas the curve for Case A shows a stronger dependence on $M_S$.
Case A can achieve exact accordance with experiment for $M_S = 2.5$TeV.
That agrees with the result in section 3.1.
For Case I we also considered
$M_S=M_{st}$ - the non-supersymmetric string.
The curve of Case I with $M_S=M_Z$ is almost degenerate with the curve
of Case I with $M_S=1$TeV.
The non-supersymmetric string beats the supersymmetric string
(Case I with $M_S=M_Z$)
with respect to gauge coupling unification.
In general, increase of $M_S$ results in a lowering of the curve for all
cases except for Case C and D.

To get an idea of the effects of higher order corrections, we perform the
calculation described
above at two-loop level for some cases.
Here, we considered Case I and II with $M_S=1$TeV
and Case I with $M_S=M_{st}$ (Figure 4).
The RGEs of the gauge couplings at two-loop level are
\eq{
{d \alpha^{-1}_a \over dt}=-{b_a \over 2 \pi}- \sum_{b=1}^3
{b_{ab} \over 8 \pi^2} \alpha_b.
}
The two-loop $\beta$-coefficients $b_{ab}$ needed in this calculation read
\cite{Jo}:
$$\bar b^{I}_{ab} = b^{SM}_{ab} = \pmatrix{199/18 & 9/2 & 44/3 \cr
                                               3/2 & 35/6 & 12 \cr
                                              11/6 & 9/2 & -26 \cr}, \quad
\bar b^{II}_{ab} = \pmatrix{307/18 & 9 & 68/3 \cr
                             3/2 & 163/6 & 12 \cr
                              11/6 & 9/2 & 22 \cr}$$
\eq{
\hbox{ and }\quad b^{MSSM}_{ab}=\pmatrix{199/9 & 9 & 88/3 \cr
                                              3 & 11 & 24 \cr
                                           11/3  & 9 & 14 \cr}.
}

In all three cases the curve is lifted by two loop corrections,
so that the accordance with the experimental data becomes worse.
The non-supersymmetric string is most sensitive
to the inclusion of higher order effects.
Case II acquires a smaller correction.
Similarly we can compute two-loop corrections for other cases,
using corresponding two-loop $\beta$-coefficients \cite{Jo}.
In these cases, two-loop corrections are as small as in Case II.

The above results show that the theoretical predictions are not compatible with
experiment for most cases.
This discrepancy might be solved, if threshold corrections due to higher
massive
modes of
string theories are taken into account.
Assuming exact
accordance at one-loop level with the measured values, we can estimate the
threshold corrections from the value of the nearest point of each curve to
the center of the error cross shown in Table 5.
This will be done in the next section.

\vspace{0.8 cm}
\leftline{\large \bf 4. String theory}
\vspace{0.8 cm}

In this section we discuss the results obtained in the previous
section from the viewpoint of orbifold models \cite{Orbi}.
The orbifold construction is one of the simplest and most promising methods to
construct four-dimensional string models.

\vspace{0.8 cm}
\leftline{\large \bf 4.1 String threshold corrections}
\vspace{0.8 cm}

String models have towers of higher massive modes, which bring about threshold
corrections to the gauge couplings.
Some parts of the string threshold corrections depend on the vacuum expectation
values of moduli fields $T$, which describe geometrical features of
orbifolds \cite{Thres}.
In general orbifolds have three independent moduli fields $T^i$ ($i=1,2,3$).
Here we restrict ourselves to the overall moduli field $T=T^i$.
The other corrections depend on the details of massive string spectra.
Large values of $T$ could lead to large threshold corrections.
The vacuum expectation value of the moduli field could be determined by a
non-perturbative superpotential $\widehat W$, which also breaks SUSY.
In refs.\cite{FILQ} the values of $O(1)$ were obtained within the framework
of the gaugino condensation scenario.
On the other hand, the values of $O(10)$ were obtained in ref.\cite{MR},
taking into account a one-loop effective potential.
Hence the gauge coupling at $M_{st}$ is written as
$$\alpha_a^{-1}=\alpha_{st}^{-1}+\Delta_a(T)+\Delta'_a,
\eqno(4.1)$$
where the second term of the right hand side is the $T$-dependent
threshold correction.
The threshold correction reads \cite{Thres},
$$\Delta_a(T)={b'_a-\delta_{GS} \over 4 \pi}\log [(T+\bar T)|\eta(T)|^4],
\eqno(4.2)$$
where $b'_a$ represents a duality anomaly coefficient and $\delta_{GS}$
denotes a Green-Schwarz coefficient \cite{GS}.
The former is determined by massless modes in string models.
Further, the Dedekind function $\eta(T)$ is expressed as
$\eta(T)=e^{-\pi T/12}\prod^\infty_{n=1}(1-e^{-2\pi nT}).$

The investigation in the previous section shows some differences between
experiment and direct string unification, except for Case~A.
Here we estimate string threshold corrections to explain the difference
between experiment and the tree-level prediction by string unification.
We represent the necessary threshold corrections by the values of $T$,
neglecting $T$-independent threshold corrections $\Delta'_a$.
In general, the bulk of the corrections comes from T-dependent parts.
In refs.\cite{Ibanez,ILR} it was discussed to explain the discrepancy
between the experiment and the string unification for the MSSM with
$M_S=M_Z$ and $k_1=5/3$, using $\Delta_a(T)$.
That analysis was extended to the cases with general values of $k_1$ in
refs.\cite{MSM}.

At first we estimate threshold corrections necessary to explain
discrepancy between the experiments and the results obtained in 3.2.
Suppose that for $\alpha_3^{-1}(M_Z)$ and $\sin^2\theta_W(M_Z)$ we denote
deviations of
the predicted values from the experiments as $\Delta \alpha_3^{-1}$ and
$\Delta \sin^2 \theta_W$, then we need the following threshold corrections
at $M_{st}$,
$$\Delta_3(T)=\Delta \alpha_3^{-1}, \quad \Delta_2(T)=\alpha^{-1}(M_Z)
\Delta \sin^2 \theta_W.
\eqno(4.3)$$
We compare the values of $\alpha_3(M_Z)$ and $\sin^2\theta_W(M_Z)$ of each
case in Table 5 with experimental values $\alpha_3(M_Z)=0.118$ and
$\sin^2\theta_W(M_Z)=0.2321$.
For example the deviations of these values for Case~I require
$\Delta_3(T)=0.344$ and $\Delta_2(T)=1.279$.
These values of the threshold corrections are realized in (4.2) by
$T=6.6$ and 19, respectively, if we take $b'_3-\delta_{GS}=1$ and
$b'_2-\delta_{GS}=1$.
These values of $T$ are found in the fifth and the sixth columns of Table 5,
where $B'_a \equiv b'_a-\delta_{GS}$.
For the other cases we can estimate necessary values of $T$ similarly.
Those values are listed in the fifth and sixth columns of Table 5.
For all of the cases, the difference $\Delta \sin^2\theta_W$ is more important
than $\Delta \alpha_3^{-1}$.
The seventh and eighth columns of Table 5 show values of $T$ deriving
necessary threshold corrections for $\Delta \sin^2\theta_W$ in the case
with $b'_2-\delta_{GS}=5$ and $b'_2-\delta_{GS}=10$, respectively.

Next we estimate the threshold corrections to explain difference between
the experiments and the results obtained in 3.1.
Including the threshold corrections we have the running gauge couplings at
$\mu$ ($\mu>M_S$) as,
$$ \alpha_a^{-1}(\mu)=\alpha_{st}-{b_a \over 2\pi}\ln{M_{st} \over \mu}
+{b'_a-\delta_{GS} \over 4 \pi}\log [(T+\bar T)|\eta(T)|^4].
\eqno(4.4)$$
Using (4.4), the string scale can be related with the unification scale
$M_X$ where $\alpha_3(M_X)=\alpha_2(M_X)$ as follows \cite{Ibanez},
$$ {\rm ln}{M_{X} \over M_{\rm st}}={\Delta b' \over 8}
 {\rm ln}[(T+\bar T)|\eta(T)|^4],\eqno(4.5)$$
where $\Delta b'\equiv b'_3-b'_2$.
Note that ${\rm ln}[(T+\bar T)|\eta(T)|^4]$ is always negative.
If $M_X<M_{St}$, the duality anomaly coefficients should satisfy
$b'_3>b'_2$.
For example the MSSM with $M_S=M_Z$ leads to $M_X=10^{16.4}$GeV.
This value of $M_X$ leads to $T=24$ in the case with $\Delta b'=1$.
Similarly we can estimate the value of $T$ for each unification scale
$M_X$ obtained in 3.1 using $M_S=1$TeV.
The results are found in the fifth, sixth and seventh columns of Table 2
for the cases with $\Delta b'=1$, 5 and 10, respectively
\footnote[4]{In ref.\cite{MSM}, possible values of $\Delta b'$ were given
explicitly for orbifold models.
To obtain $\Delta b'=5$ or 10 gives severe constraints
on model building.}.
The numbers in the parentheses of Table 2 correspond to $T$ in the
case with $M_S=\sqrt {10}$TeV.

\vspace{0.8 cm}
\leftline{\large \bf 4.2 Level of $U(1)_Y$}
\vspace{0.8 cm}

Here we comment on the value of $k_1$.
The discussion in 3.2 seems to show that the desirable value is $k_1=1.3
- 1.4$ for the string unification of the MSSM.
Within the framework of orbifold models massless states satisfy the
following condition \cite{Orbi},
$$h+N_{OSC}+c-1=0,
\eqno(4.6)$$
where $N_{OSC}$ is the oscillator number, $c$ is the ground state energy
and $h$ is the conformal dimension due to the gauge parts.
A state belonging to the representation $\underline{R}$ of a non-abelian group
$G$
contributes to the conformal dimension as follows,
$$h={C(\underline {R}) \over C(G)+k}.
\eqno(4.7)$$
A state transforming under an abelian group with charge $Q$ gives
$h=Q^2/k_1$.
Under the condition that the MSSM matter fields are massless, the level of
$U(1)_Y$ is restricted by a lower bound.
These lower bounds are shown explicitly in refs.\cite{MSM}.
For example the twisted sector of the $Z_3$ orbifold has $c=1/3$.
The level should satisfy the condition $k_1 \geq 4/3$ so that the chiral
field $U$ appears in the twisted sector.
Further the existence of the chiral field $E$ in the twisted sector
requires $k \geq 3/2$, although the other MSSM matter fields in the
twisted sector are allowed in the case with $k_1 \geq 1$.
However, the condition on $k_1$ restricts not only $U$ and $E$ fields, but
also the others, taking into account Yukawa couplings.
In $Z_3$ orbifold models, the field $E$ in the twisted sector is ruled out and
the twisted sector
cannot contain $U$ for some cases.
Orbifold models have selection rules for Yukawa couplings different from those
obtained
by gauge invariance \cite{Yukawa}.
These selection rules restrict the couplings of one sector to another
ref.\cite{KO}.
In $Z_3$ orbifold models, the untwisted matter fields are allowed
to couple to the untwisted matter fields only.
If the fields $U$ and $E$ are permitted in the untwisted sector only,
the fields $Q$ and $L$ and the Higgs fields are allowed only in the
untwisted sector to give the Yukawa couplings.
Further this fact has a phenomenological implication that these
Yukawa couplings give the same magnitude, i.e. Yukawa coupling
unification, although twisted sectors could lead to a hierarchy
structure of Yukawa couplings \cite{hiera}.
For the other orbifold models, we can discuss similar constraints.
In general the field $E$ is restricted to the twisted sector.
Therefore the prediction of $k_1$ has sensitive effects on model building.

\vspace{0.8 cm}
\leftline{\large \bf 5. Conclusion}
\vspace{0.8 cm}

We have studied the gauge coupling unification in cases with
non-universal soft scalar and gaugino masses.
The unification scale of $SU(3)$ and $SU(2)$ gauge couplings is
sensitive to the non-universality.
Some cases lead to a unification scale $M_X$ around $M_{st}$.
The ratio $\alpha_x/\alpha_1$ at $M_X$ is also sensitive to
the non-universality and varies from 1.4--1.7 for $M_S=1$TeV.
We have also run the gauge couplings top-down assuming the string
unification of the MSSM.
That analysis seems to show that the preferred values of $k_1$ are in
the range $1.3 - 1.4$.
These values of $k_1$ give some constraints on model building.
Two-loop corrections depend on the non-universality, too.
The non-SUSY case acquires larger two-loop corrections.
The corrections are reduced by effects of some superpartners.

Several cases show discrepancies between experiment and the predictions by
string unification of the MSSM.
These discrepancies could be explained in terms of threshold
corrections due to massive string modes.
Threshold corrections are expressed in terms of the vacuum expectation values
of the moduli fields.
Some cases require $T$ of order $O(10)$.

It is interesting to analyze the other RGEs with decoupling as
discussed here and
to study the radiative symmetry.
We have assumed that the SUSY breaking scale for the Higgs sector
is $M_S$.
One has to investigate the condition of the successful radiative
symmetry breaking in the mass spectra of the non-universal cases.
Further it is important to investigate which phenomenological
aspects are sensitive or insensitive to the non-universality of
 other soft SUSY breaking parameters as well as
soft scalar masses and gaugino masses.

\vspace{0.8 cm}
\leftline{\large \bf Acknowledgement}
\vspace{0.8 cm}

The authors would like to thank J.~Louis for useful discussions.
They also acknowledge to M.~Konmura, N.~Polonsky, D.~Suematsu, K.~Yamada
and Y.~Yamagishi.


\newpage
\footnotesize
\pagestyle{empty}
{\large Table 1: Non-universal cases}

\begin{tabular}{|c|c|c|c|}\hline
Case & ${\cal A}$     & ${\cal B}$ &
$\bar b_1,\bar b_2, \bar b_3$ \\ \hline \hline
 I   & $Q,~U,~D,~L,~E,~\lambda_3,~\lambda_2,~\lambda_1$ &
& $41/6,-19/6,-7$ \\
 II  & $Q,~U,~D,~L$   & $E,~\lambda_3,~\lambda_2,~\lambda_1$
& $47/6,-11/6,-5$ \\
 III & $Q,~L$         & $U,~D,~E,~\lambda_3,~\lambda_2,~\lambda_1$
& $19/2,-11/6,-4$ \\
 IV  & $L$            & $Q,~U,~D,~E,~\lambda_3,~\lambda_2,~\lambda_1$
& $29/3,-1/3,-3$ \\
 V   & $Q,~U,~D$      & $L,~E,~\lambda_3,~\lambda_2,~\lambda_1$
& $25/3,-4/3,-5$ \\
 VI  & $U,~D$         & $Q,~L,~E,~\lambda_3,~\lambda_2,~\lambda_1$
& $17/2,1/6,-4$ \\
 A   & $Q,~L,~\lambda_2$ & $U,~D,~E,~\lambda_3,~\lambda_1$
& $19/2,-19/6,-4$ \\
 B   & $Q,~U,~D,~L,~\lambda_2$ & $E,~\lambda_3,~\lambda_1$
& $47/6,-19/6,-5$ \\
 C   & $Q,~U,~D,~\lambda_3$ & $L,~E,~\lambda_2,~\lambda_1$
& $25/3,-4/3,-7$ \\
 D   & $Q,~U,~D,~L,~\lambda_3$ & $E,~\lambda_2,~\lambda_1$
& $47/6,-11/6,-7$ \\ \hline
\end{tabular}

\vskip 0.5cm

{\large Table 2: Unification scale}

\begin{tabular}{|c|c|c|c|c|c|c|}\hline
Case & $\log_{10}M_X$ & $\alpha_X^{-1}$ & $\alpha_X/\alpha_1(M_X)$ &
$T(\Delta b'=1)$ & $T(\Delta b'=5)$ & $T(\Delta b'=10)$ \\
\hline \hline
I  & 16.5   & 26.0   & 1.59   & 23   & 6.2   & 3.8   \\
     & (16.5) & (26.7) & (1.57) & (23) & (6.2) & (3.8) \\
II & 16.7   & 25.6   & 1.58   & 19   & 5.4   & 3.4   \\
     & (16.8) & (26.3) & (1.55) & (17) & (4.9) & (3.1) \\
III& 16.9   & 25.3   & 1.52   & 15   & 4.5   & 2.8   \\
     & (17.1) & (25.7) & (1.47) & (11) & (3.5) & (2.3) \\
IV & 16.8   & 24.8   & 1.57   & 17   & 4.9   & 3.1   \\
     & (16.9) & (25.0) & (1.54) & (15) & (4.5) & (2.8) \\
V  & 16.5   & 25.2   & 1.60   & 23   & 6.2   & 3.8   \\
     & (16.6) & (25.7) & (1.59) & (21) & (5.8) & (3.5) \\
VI & 16.4   & 24.8   & 1.63   & 24   & 6.6   & 4.1   \\
     & (16.4) & (25.0) & (1.64) & (24) & (6.6) & (4.1) \\
A  & 17.3   & 25.7   & 1.44   & 7.3  & 2.5   & 1.6   \\
     & (17.7) & (26.3) & (1.31) & --   & --    & --    \\
B  & 17.0   & 25.8   & 1.50   & 13   & 4.0   & 2.6   \\
     & (17.3) & (26.4) & (1.44) & (7.3)& (2.5) & (1.6) \\
C  & 16.0   & 25.4   & 1.67   & 32   & 8.2   & 5.0   \\
     & (15.8) & (25.9) & (1.69) & (35) & (9.0) & (5.4) \\
D  & 16.1   & 25.6   & 1.65   & 30   & 7.8   & 4.7   \\
     & (16.0) & (26.2) & (1.66) & (32) & (8.2) & (5.0) \\ \hline
\end{tabular}

\newpage

{\large Table 3: Radiative corrections to gaugino masses}

\begin{tabular}{|c|c|c|c|c|c|}\hline
Case & I & A & B   & C  & D \\ \hline \hline
$M_3/M_3(M_X)$ &  2.3  &  3.3  &  3.6  &  2.3  &  2.3  \\
               & (2.2) & (3.3) & (3.9) & (2.1) & (2.1) \\ \hline
$M_2/M_2(M_X)$ &  0.84  &  0.83  &  0.83  &  0.89  &  0.91  \\
               & (0.85) & (0.84) & (0.84) & (0.93) & (0.95) \\ \hline
$M_1/M_1(M_X)$ &  0.43  &  0.38  &  0.40  &  0.44  &  0.44  \\
               & (0.44) & (0.37) & (0.40) & (0.45) & (0.45) \\ \hline
\end{tabular}

\vskip .5cm
{\large Table 4: Radiative corrections to soft scalar masses}

\begin{tabular}{|c|ccc|ccc|}\hline
Case & \multicolumn{3}{c|} {A} & \multicolumn{3}{c|} {B} \\ \cline{2-7}
\hline \hline
 & $M_3^2$ & $M_2^2$ & $M_1^2$ & $M_3^2$ & $M_2^2$ & $M_1^2$ \\ \hline
$\Delta m_Q^2$ & 5.0 & 0.46 & $4.3\times 10^{-3}$ & 4.6 & 0.46
& $4.2 \times 10^{-3}$ \\
& (4.5) & (0.46) & $(4.3 \times 10^{-3})$ & (4.0) & (0.44) &
 $(4.2 \times 10^{-3})$ \\ \hline
$\Delta m_U^2$ & 6.7 & --- & $7.0\times 10^{-2}$ & 4.6 & ---
& $6.8 \times 10^{-2}$ \\
& (6.7) & (---) & $(7.1 \times 10^{-2})$ & (4.0) & (---) &
 $(6.8 \times 10^{-2})$ \\ \hline
$\Delta m_L^2$ & --- & 0.46 & $3.9\times 10^{-2}$ & --- & 0.46
& $3.8 \times 10^{-2}$ \\
& (---) & (0.46) & $(3.9 \times 10^{-2})$ & (---) & (0.44) &
 $(3.8 \times 10^{-2})$ \\ \hline
$\Delta m_E^2$ & --- & --- & $1.6\times 10^{-1}$ & --- & ---
& $1.5 \times 10^{-1}$ \\
& (---) & (---) & $(1.6 \times 10^{-1})$ & (---) & (---) &
 $(1.6 \times 10^{-1})$ \\ \hline
\end{tabular}

\begin{tabular}{|c|ccc|ccc|} \hline
Case & \multicolumn{3}{c|} {C} & \multicolumn{3}{c|} {D} \\ \cline{2-7}
\hline \hline
 & $M_3^2$ & $M_2^2$ & $M_1^2$ & $M_3^2$ & $M_2^2$ & $M_1^2$ \\ \hline
$\Delta m_Q^2$ & 3.8 & 0.44 & $4.0\times 10^{-3}$ & 3.8 & 0.44
& $4.0 \times 10^{-3}$ \\
& (3.0) & (0.41) & $(3.9 \times 10^{-3})$ & (3.1) & (0.41) &
 $(4.0 \times 10^{-3})$ \\ \hline
$\Delta m_U^2$ & 3.8 & --- & $6.5\times 10^{-2}$ & 3.8 & ---
& $6.5 \times 10^{-2}$ \\
& (3.0) & (---) & $(6.3 \times 10^{-2})$ & (3.1) & (---) &
 $(6.4 \times 10^{-2})$ \\ \hline
$\Delta m_L^2$ & --- & 0.45 & $3.7\times 10^{-2}$ & --- & 0.44
& $3.7 \times 10^{-2}$ \\
& (---) & (0.43) & $(3.7 \times 10^{-2})$ & (---) & (0.41) &
 $(3.6 \times 10^{-2})$ \\ \hline
$\Delta m_E^2$ & --- & --- & $1.5\times 10^{-1}$ & --- & ---
& $1.5 \times 10^{-1}$ \\
& (---) & (---) & $(1.5 \times 10^{-1})$ & (---) & (---) &
 $(1.5 \times 10^{-1})$ \\ \hline
\end{tabular}
\newpage
{\large Table 5: Nearest points to experimental value}

\begin{tabular}{|l|c|c|c|c|c|c|c|}
\hline
Case & $\alpha_3$ & $\sin^2 \theta_W$ & $k_1$
& $T(B'_3=1)$ &$T(B'_2=1)$ &$T(B'_2=5)$ &$T(B'_2=10)$  \\
\hline\hline
 I & 0.123 & 0.242 & 1.34 & 6.6 & 19 & 5.3 & 3.4 \\
 II & 0.123 & 0.240 & 1.37 & 6.6 & 16 & 4.6 & 2.9 \\
 III & 0.121 & 0.238 & 1.37 & 4.7 & 12 & 3.8 & 2.4 \\
 IV & 0.122 & 0.239 & 1.39 & 5.7 & 14 & 4.2 & 2.7 \\
 V & 0.123 & 0.241 & 1.36 & 6.6 & 17 & 5.0 & 3.1 \\
 VI & 0.123 & 0.242 & 1.37 & 6.6 & 19 & 5.3 & 3.4 \\
 A & 0.120 & 0.235 & 1.38 & 3.6 & 7.2& 2.4 & 1.5 \\
 B & 0.121 & 0.237 & 1.38 & 4.7 & 11 & 3.4 & 2.2 \\
 C & 0.126 & 0.246 & 1.31 & 9.2 & 25 & 6.8 & 4.2 \\
 D & 0.125 & 0.245 & 1.32 & 8.4 & 24 & 6.4 & 4.0 \\
\hline
\end{tabular}
\vfill
\normalsize
{\large\bf Figure captions:}
\begin{enumerate}
\item Unification scale $M_X$ as a function of $M_S$ - the SUSY breaking scale
-
for
      Case I and Case A - D, where x and y denote $\log_{10}M_S$ and
$\log_{10}M_X$
      respectively.
\item One-loop calculation of $\sin^2\theta_W(\alpha_3)$ at $M_Z$ with $M_S =
1$TeV
      for all the cases under consideration, i. e. Case I - VI and Case A - D.
In this and
      the following figures a stands for $\alpha_3(M_Z)$ and sin for
$\sin^2\theta_W(M_Z)$.
\item One-loop calculation of $\sin^2\theta_W(M_Z)$ at $M_Z$ for Case I, II and
A for
      different values of $M_S$.
\item One- and two-loop curves of $\sin^2\theta_W(\alpha_3)$ at $M_Z$ with $M_S
= 1$TeV
      for Case I and II and with $M_S = M_{st}$ for Case I (non-supersymmetric
string).
\end{enumerate}


\newpage

\begin{thebibliography}{99}

\bibitem{Nilles}
For a review, see e.g. H.-P.~Nilles, Phys.~Rep. {\bf 110} (1984) 1.

\bibitem{Ibanez}
L.E.~Ib\'a\~nez and D.~L\"ust, Nucl.~Phys. {\bf B382} (1992) 305.

\bibitem{KL}V.~S.~Kaplunovsky and J.~Louis,
 Phys.~Lett. {\bf B306} (1993) 269.

\bibitem{Brignole}
A.~Brignole, L.~E.~Ib\' a\~ nez and C.~Mu\~ noz, Nucl.~Phys. {\bf B422}
(1994) 125.

\bibitem{KSYY}
T.~Kobayashi, D.~Suematsu, K.~Yamada and Y.~Yamagishi,
preprint Kanazawa-94-16 (hep-ph/9408322) to be published in Phys.~Lett.~B.

\bibitem{KSY}
T.~Kobayashi, D.~Suematsu and Y.~Yamagishi, Phys.~Lett. {\bf B329} (1994) 27.

\bibitem{Non}
A.~Lleyda and C.~Mu\~noz, Phys.~Lett. {\bf B317} (1993) 82.

N.~Polonsky and A.~Pomarol, Phys.~Rev.~Lett. {\bf 73} (1994) 2292;
preprint UPR-0627T (hep-ph/9410231).

Y.~Kawamura, H.~Murayama and M.~Yamaguchi, preprint DPSU-9402 \\
(hep-ph/9406245).

M.~Carena and C.~Wagner, preprint CERN-TH-7321/94 (hep-ph/9407209).

D.~Matalliotakis and H.P.~Nilles, Nucl.~Phys. {\bf B435} (1995) 115.

M.~Olechowski and S.~Pokorski, Phys.~Lett. {\bf B344} (1995) 201.

T.~Kobayashi, M.~Konmura, D.~Suematsu, K.~Yamada and Y.~Yamagishi,
preprint Kanazawa-94-17(hep-ph/9410269).

Ph.~Brax, U.~Ellwanger and C.A.~Savoy, preprint LPTH Orsay 94/101
(hep-ph-9411397).

\bibitem{FCNC}
For recent work, see

J.~Hagelin, S.~Kelly and T.~Tanaka, Mod.~Phys.~Lett. {\bf A8} (1993) 2737;
 Nucl.~Phys. {\bf B415} (1994) 293.

D.~Choudhury, F.~Eberlein, A.~K\"onig, J.~Louis and S.~Pokorski,
Phys.~Lett. {\bf B342} (1995) 180.

J.~Louis and Y.~Nir, preprint LMU-TPW 94-17 (hep-ph/94114229).

\bibitem{MX}
J.~Ellis, S.~Kelley and D.V.~Nanopoulous, Phys.~Lett. {\bf B260} (1991) 131.

U.~Amaldi, W.~de~Boer and H.~F\"urstenau, Phys.~Lett. {\bf B260} (1991) 447.

P.~Langacker and M.~Luo, Phys.~Rev. {\bf D44} (1991) 817.

G.G.~Ross and R.G.~Roberts, Nucl.~Phys. {\bf B377} (1992) 571.

P.~Langacker and N.~Polonsky, Phys.~Rev. {\bf D47}(1993)4028.

\bibitem{Kaplunovsky}
V.S.~Kaplunovsky, Nucl.~Phys. {\bf B307} (1988) 145.

\bibitem{Gi}
P.~Ginsparg, Phys.~Lett. {\bf B197} (1987) 139.

\bibitem{Fo}
A.~Font, L.E.~Ib\'a\~nez and F.~Quevedo, Nucl. Phys. {\bf B345} (1990) 389.

\bibitem{highk}
D.~Lewellen, Nucl.~Phys. {\bf B337} (1990) 61.

G.~Cleaver, preprint OHSTPY-HEP-T-94-007 (hep-th/9409096).

G.~Aldazabal, A.~Font, L.E.~Ib\'a\~nez and A.~Uranga, preprint
FTUAM-94-28 (hep-th/9410206).

S.~Chaudhuri, S.-w.~Chung, G.~Hockney and J.~Lykken, preprint
FERMILAB-PUB-94/413-T (hep-ph/9501361).

\bibitem{Ib}
L.E.~Ib\'a\~nez, Phys.~Lett. {\bf B318} (1993) 73.

\bibitem{Orbi}
L.~Dixon, J.~Harvey, C.~Vafa and E.~Witten, Nucl.~Phys. {\bf B261} (1985)
 678; Nucl.~Phys. {\bf B274} (1986) 285.

L.E.~Ib\'a\~nez, J.~Mas, H.P.~Nilles and F.~Quevedo, Nucl.~Phys. {\bf B301}
 (1988) 157.

Y.~Katsuki, Y.~Kawamura, T.~Kobayashi, N.~Ohtsubo, Y.~Ono and \\
K.~Tanioka, Nucl.~Phys. {\bf B341} (1990) 611.

\bibitem{Kahler}
E.~Witten, Phys.~Lett. {\bf B155} (1985) 151.

S.~Ferrara, C.~Kounnas and M.~Porrati, Phys.~Lett. {\bf B181} (1986) 263.

M.~Cveti\u{c}, J.~Louis and B.~Ovrut, Phys.~Lett. {\bf B206} (1988) 227.

L.J.~Dixon, V.S.~Kaplunovsky and J.~Louis, Nucl.~Phys. {\bf B329} (1990) 27.

\bibitem{Thres}
L.J.~Dixon, V.S.~Kaplunovsky and J.~Louis, Nucl.~Phys. {\bf B355} (1991) 649.

I.~Antoniadis, K.S.~Narain and T.R.~Taylor, Phys.~Lett. {\bf B267} (1991) 37.

J.-P.~Derendinger, S.~Ferrara, C.~Kounnas and F.~Zwirner, Nucl.~Phys.
 {\bf B372} (1992) 145.

\bibitem{BLM}
R.~Barbieri, J.~Louis and M.~Moretti, Phys.~Lett. {\bf B312} (1993) 451.

\bibitem{Carlos}
B.~de~Carlos, J.~A.~Casas and C.~Mu\~noz, Phys.~Lett. {\bf B299} (1993) 234.

\bibitem{Chankowski}
P.H.~Chankowski, Phys.~Rev. {\bf D41} (1990) 2877.


\bibitem{Jo}
D.R.T.~Jones, Phys.~Rev. {\bf D25} (1982) 581.

\bibitem{FILQ}
A.~Font, L.~E.~Ib\'a\~nez, D.~L\"ust and F.~Quevedo,
Phys.~Lett. {\bf B245}(1990)401.

S.~Ferrara, N.~Magnoli, T.~R.~Taylor and G.~Veneziano, Phys.~Lett.
{\bf B245}(1990)409.

M.~Cveti\u{c}, A.~Font, L.~E.~Ib\'a\~nez, D.~L\"ust and F.~Quevedo,
Nucl.~Phys. {\bf B361} (1991) 194.

B.~de~Carlos, J.~A.~Casas and C.~Mu\~noz, Nucl.~Phys. {\bf B399}(1993)623.

\bibitem{MR}
A.~de~la~Macorra and G.~G.~Ross, Nucl.~Phys. {\bf B404}(1993)145;
Phys.~Lett. {\bf 325} (1994) 85.

\bibitem{GS}
M.B.~Green and J.H.~Schwarz, Phys.~Lett. {\bf B149} (1984) 117.

J.-P.~Derendinger, S.~Ferrara, C.~Kounnas and F.~Zwirner,
 Phys. Lett. {\bf B271} (1991) 307.

\bibitem{ILR}
L.~E.~Ib\' a\~ nez, D.~L\" ust and G.~G.~Ross, Phys.~Lett.
{\bf B272} (1991) 251.

\bibitem{MSM}
H.~Kawabe, T.~Kobayashi and N.~Ohtsubo, Phys.~Lett. {\bf B322} (1994) 331;
Phys.~Lett. {\bf B325} (1994) 77; Nucl.~Phys. {\bf B434} (1995) 210.

T.~Kobayashi, Phys.~Lett. {\bf B326} (1994) 231; preprint Kanazawa-94-10
(hep-ph/9406238) to be publishied in Int.J.~Mod.~Phys. A.

\bibitem{Yukawa}
L.~Dixon, D.~Friedan, E.~Martinec and S.~Shenker, Ncul.~Phys.
{\bf B282} (1987) 13.

Hamidy and C.~Vafa, Nucl.~Phys. {\bf B279} (1987) 465.

\bibitem{KO}
T.~Kobayashi and N.~Ohtsubo, Phys.~Lett. {\bf B245} (1990) 441;
Phys.~Lett. {\bf B262} (1991) 425; Int.J.~Mod.~Phys.~{\bf A9} (1994) 87.


\bibitem{hiera}
L.E.~Ib\'a\~nez, Phys.~Lett. {\bf B181} (1986) 269.

J.A~Casas and C.~Mu\~noz, Nucl.~Phys. {\bf B332} (1990) 189.

J.A~Casas, F.~Gomez and C.~Mu\~noz, Phys.~Lett. {\bf B292} (1992) 42.


\end{thebibliography}
\end{document}